\newcommand{\beq}{\begin{equation}}
\newcommand{\eeq}{\end{equation}}
\newcommand{\beqn}{\begin{eqnarray}}
\newcommand{\eeqn}{\end{eqnarray}}
\newcommand{\epem}{$\mbox{e}^+ \mbox{e}^-$}

\newcommand{\nonu}{\nonumber\\}
\newcommand{\porder}[1]{\mbox{${\cal O}(#1)$}}

\newcommand{\SH}{{S_{\!\scriptscriptstyle H}}}
\newcommand{\XB}{{x_{\!\scriptscriptstyle B}}}
\newcommand{\PP}{{P_{\!\scriptscriptstyle 0}}}
\newcommand{\EE}{{E_{\scriptscriptstyle 0}}}
\newcommand{\EH}{{E_{\mbox{\scriptsize h}}}}
\newcommand{\zh}{{z_{\mbox{\scriptsize h}}}}

\newcommand{\AAA}{\mbox{\scriptsize A}}
\newcommand{\hh}{\mbox{\scriptsize h}}
\newcommand{\dPS}{\mbox{dPS}}
\newcommand{\dd}{\mbox{d}}
\newcommand{\Pii}{P}
\newcommand{\qb}{\overline{q}}

\newlength{\conserve}
\newlength{\wideformula}
\documentstyle[11pt,fleqn]{article}

\newlength{\dinwidth}
\newlength{\dinmargin}
\begin{document}
\begin{titlepage}

\renewcommand{\thefootnote}{\fnsymbol{footnote}}
\hfill
\hspace*{\fill}
\begin{minipage}[t]{5cm}
CERN-TH.7300/94\\
hep-ph/9406274
\end{minipage}
\begin{center}

\vspace{8mm}
{\LARGE
One-Particle Inclusive Processes\\
in Deeply Inelastic Lepton--Nucleon Scattering\\
}
\vspace{1cm}
{\Large
Dirk Graudenz$\;$\footnote[3]
{{\em Electronic mail address: 
graudenz@cernvm.cern.ch}}\\
Theoretical Physics Division, CERN\\
CH-1211 Geneva 23\\
}
\end{center}
\vspace{0.5cm}
%\begin{center}
%\fbox{\fbox{D R A F T}}
%\end{center}

\vspace{1cm}

\begin{abstract}
\noindent
The one-particle inclusive cross section in deeply inelastic
lepton--nucleon scattering, expressed in terms of parton densities
and fragmentation functions being differential in the invariant mass
of the observed hadron and of the incoming nucleon, diverges if this
invariant mass vanishes. This divergence can be traced back to the
kinematical configuration where the parent parton of the observed hadron
is emitted collinearly from the incoming parton of the QCD subprocess.
By using the concept of 
``fracture functions'', which has recently been introduced by
Trentadue and Veneziano, it is possible to 
absorb this divergence in these new distribution functions
as long as the observed hadron is not soft.
This procedure allows the determination of a finite one-particle
inclusive cross section in next-to-leading order QCD perturbation theory.
We give details of the calculation and the explicit form of
the bare fracture functions in terms of the renormalized ones.
\end{abstract}

\vspace{8cm}
\noindent
\begin{minipage}[t]{5cm}
CERN-TH.7300/94\\
June 1994
\end{minipage}
\end{titlepage}

\section{Introduction}
\noindent
QCD as a theory of the strong interactions should be able to give a description
of one-particle inclusive processes. Since non-perturbative effects
such as the nucleon wave functions and the 
fragmentation of quarks and gluons
are not manageable in explicit 
calculations, it is necessary to parametrize them
by means of phenomenological distribution functions in order to make
contact with perturbative QCD 
\cite{1,2,3,4,5,6,7}.
The case of one-particle inclusive
lepton--nucleon scattering has been considered by several authors
\cite{8,9,5,6,10,11,12,13}. 
The canonical procedure to do the calculation is to define a variable
$\zh=\PP h/\PP q$ as in \cite{5}, where $\PP$ is the momentum of the 
incoming nucleon, $q$ is the momentum of the exchanged virtual photon, 
and $h$ is the momentum of the observed hadron.
It turns out that $\dd \sigma/\dd \zh$,
the differential cross section
for the production of h plus anything else,
is singular for $\zh\rightarrow 0$.
A similar effect ($\dd \sigma/\dd \zh\rightarrow \infty$ for $\zh\rightarrow 0$)
can be observed in the case of \epem annihilation, where 
$\zh \rightarrow 0$ means that the observed hadron h is soft 
($\EH\rightarrow 0$). 
This singularity is physical, being related
to the fact that the multiplicity for the emission of
soft particles is infinite.
However, in lepton--nucleon scattering the situation is different:
$\zh \rightarrow 0$ means (in the case of massless particles) that either
$\EH\rightarrow 0$ or that h is collinear with the target remnant.
In the case $\EH\rightarrow 0$ the same argument as in the \epem case applies.
If, however, $\zh\rightarrow 0$, but not $\EH\rightarrow 0$,
the process under consideration is that of a hadron h with
non-vanishing energy emitted in the remnant direction, which is
physically perfectly well defined and has to be distinguished 
from the case of a soft process.
It is possible to separate the two cases by introducing energy and
angle variables $z$ and $v$ ($z=0$ means $\EH=0$, 
$v=1$ means that h is collinear with the remnant) for the observed hadron.
This allows to consider the case of $\zh\rightarrow 0$ without
having $\EH\rightarrow 0$. 
$\dd\sigma/\dd z \dd v$ can be calculated in perturbative QCD by means
of parton densities and fragmentation functions.
Again, it turns out that $\dd\sigma/\dd z \dd v \rightarrow \infty$
for $z \not= 0,\, v\rightarrow 1$. A closer analysis of the problem reveals
that it stems from the contributions in which the fragmentation function
is attached to an outgoing parton which is emitted
from the incoming parton. It is clear that 
a collinear divergence appears if the mentioned outgoing parton
is collinear to the incoming one.
In this phase-space region, the observed hadron h comes
from the target remnant region.
This singularity cannot be absorbed into the
parton distribution functions or the fragmentation functions since in lowest
order (in the formalism based solely on parton densities and 
fragmentation functions) the observed hadron always originates in the current
region.
To absorb the divergence, it is necessary to introduce a distribution
function that describes hadrons in the target remnant region.
This has been done in \cite{14}; the new distribution functions
have been called ``fracture functions''.
In this paper it is shown that, using these concepts, a finite cross section
$\dd\sigma/\dd z \dd v$ for deeply inelastic lepton--nucleon scattering
can be calculated for $\EH \not= 0$ in next-to-leading order QCD
perturbation theory.

\noindent
The outline of this paper is as follows. In Section~\ref{sect2}
the calculation of one-particle inclusive cross sections
is briefly reviewed.
The \porder{\alpha_s} QCD corrections
to the parton-model process are determined in Section~\ref{sect3}.
In Section~\ref{sect4} the explicit form of the renormalized fracture
function in terms of bare distribution functions is stated.
The finite cross section is discussed in Section~\ref{finite}.
The paper closes with a summary and some conclusions.
Technical details and explicit formulas are relegated to appendices.

\newpage 
\section{One-Particle Inclusive Cross Sections}
\label{sect2}
\noindent
In the following the scattering process 
\beq
   \mbox{l}+\mbox{A} \rightarrow 
   \mbox{l$^\prime$}+\mbox{h}+\mbox{X}
\eeq
is considered,
where l and A are the incoming charged lepton and nucleon,
respectively, l$^\prime$ is the scattered charged lepton, h is the identified
hadron, and X denotes anything else in the hadronic final state.
Let $l$ and $l^\prime$ be the momenta of the incoming and outgoing lepton,
respectively, $\PP$ the momentum of the incoming nucleon and $h$ the
momentum of the outgoing identified hadron. The integration over the angles
that determine the relative orientation of the leptonic and hadronic
final state can be performed. 
The remaining lepton phase-space variables are the Bjorken
variables
\beq
\XB=\frac{Q^2}{2\PP q}, \quad y=\frac{\PP q}{\PP l}.
\eeq
Here $q=l-l^\prime$ 
is the momentum transfer and $Q^2=-q^2>0$ is the virtuality of
the exchanged virtual photon.

\noindent
The cross section for the production of an $n$-parton final state 
differential in $\XB$, $y$ and the phase-space of the final-state
partons is
\setlength{\conserve}{\mathindent}
\setlength{\mathindent}{\wideformula}
\beqn
\quad
\frac{\mbox{d}\sigma}{\mbox{d}\XB\mbox{d}y\,\mbox{dPS}^{(n)}}&=&
\sum_i\int\frac{\mbox{d}\xi}{\xi}\,P_{i/\AAA}(\xi)\,\,\alpha^2 
\,\frac{1}{2\SH\XB}\,\frac{1}{e^2(2\pi)^{2d}}
\,\,\left(Y^M\,\left(-g^{\mu\nu}\right)
            +Y^L\,
             \frac{4\XB^2}{Q^2}\,\PP^\mu\PP^\nu
       \right) H_{\mu\nu}.
\eeqn
\setlength{\mathindent}{\conserve}
Here the case of one-photon exchange has been considered; 
the exchange of weak vector bosons is neglected for simplicity.
$\SH=(\PP+l)^2$ is the square of the total CM energy of the
lepton--nucleon scattering process, $\xi$ is the momentum fraction of
the initial parton of the QCD subprocess, and $P_{i/\AAA}(\xi)$ is the
probability distribution function for parton $i$ in the nucleon A; 
$P_{i/\AAA}(\xi)$ can be either a parton density $f_{i/\AAA}(\xi)$ or
a fracture function $M_{i,h/\AAA}(\xi,\zeta)$, see below.
$\alpha=e^2/4\pi$ is the fine structure constant and
$d=4-2\epsilon$ is the space-time dimension used in dimensional
regularization.
A factor of $1/4$ for the average over the
spin degrees of freedom of the incoming particles
is already included.
The last factor, $H_{\mu\nu}$, is the hadron tensor
defined by 
\beq
H_{\mu\nu}=\sum_{\mbox{\scriptsize spins}}
{\overline{{\cal M}_\mu}\,{\cal M}_\nu},
\eeq
where $\epsilon^\mu(\lambda){\cal M}_\mu$ is the matrix element for the process 
\beq
   \gamma^*+\mbox{parton} \rightarrow 
   n\,\,\mbox{partons},
\eeq
with $\epsilon^\mu(\lambda)$ the polarization vector of 
a virtual photon with polarization $\lambda$. The ratios
\beq
Y^M=\frac{1+(1-y)^2}{2y^2} \quad\mbox{and}\quad
Y^L=\frac{4(1-y)+1+(1-y)^2}{2y^2}
\eeq
specify the $y$-dependence of the contributions from the 
two photon polarizations.
The projections operating on 
$H_{\mu\nu}$ are the result of the integration over the angles
that describe the lepton-hadron orientation.
The cross section consists of two parts, proportional to 
$Y^M$, the ``metric'' contribution
(it is obtained by a contraction of the hadron tensor with
$(-g^{\mu\nu})$), and proportional to 
$Y^L$, the longitudinal contribution.

\noindent
For the production of an identified hadron h in the final state, there are
in principle two mechanisms \cite{15}: 
h may originate from the current (in the
parton-model, from the produced quark, or in the QCD-improved parton-model,
from one of the produced partons of the QCD subprocess) or from the
target remnant \cite{14}, so that
\beq
\sigma\,=\,\sigma_{\mbox{\scriptsize current}}\,+
\sigma_{\mbox{\scriptsize target}}.
\eeq
These two contributions cannot be disentangled
in the QCD-improved parton-model, because
$\sigma_{\mbox{\scriptsize current}}$ gives a contribution to
$\sigma_{\mbox{\scriptsize target}}$ 
in the case of collinear singularities in the
initial state. This fact is reflected
in the definition of the renormalized fracture functions, which 
are, contrary to the case of parton densities, inhomogeneous
as a function of the bare fracture functions, as can be 
seen in Section~\ref{renden}.
In the QCD-improved parton-model, $\sigma_{\mbox{\scriptsize current}}$ 
and
$\sigma_{\mbox{\scriptsize target}}$ are 
given by the graphs in Figs.~1~(a) and 1~(b),
respectively; $\sigma_{\mbox{\scriptsize current}}$ 
can be described by means of
parton densities $f_{i/\AAA}(\xi)$ and fragmentation functions
$D_{h/i}(z)$. 
The lowest-order contribution is given by the graph in Fig.~2~(a).
$\sigma_{\mbox{\scriptsize target}}$, 
however, needs a new phenomenological input, 
the ``fracture function'' $M_{i,h/\AAA}(\xi,\zeta)$, giving the probability
to find a parton $i$ with momentum fraction $\xi$ and a hadron h
with momentum fraction $\zeta$ in the nucleon A, 
for the parton model process see Fig.~2~(b).
The kinematical restriction on $\xi$ and $\zeta$ is
$\xi+\zeta\leq 1$.

\noindent
It is reasonable to 
consider the production process in the CM frame of the incoming nucleon and
the incoming virtual photon, so $\vec{\PP}+\vec{q}=\vec{0}$.
The positive $z$-axis is defined by the $q$-direction.
The hadron~h has polar angle $\vartheta$ 
relative to the virtual photon and energy $E_{\hh}$ in this frame.
The energy of the incoming nucleon is
\beq
\EE\,=\,\frac{Q}{2}\,\frac{1}{\sqrt{\XB(1-\XB)}}.
\eeq
Two new variables $v$, $z$ can be defined 
by\footnote{The variable $z$ has no relation to the variable $\zh$
mentioned in the introduction.}
\beq
v=\frac{1}{2}(1-\cos\vartheta), \quad z=\frac{\EH}{\EE(1-\XB)}.
\eeq
The range of these variables is $v,\,z\in[0,1]$
if the masses of all particles are neglected.
The differential cross section calculated in the following is
\beq
\frac{\mbox{d}\sigma(\mbox{l}+\mbox{A} \rightarrow 
   \mbox{l$^\prime$}+\mbox{h}+\mbox{X}
)}{\mbox{d}\XB\,\mbox{d}y\,\mbox{d}z\,\mbox{d}v}.
\eeq
It turns out that QCD-corrections to the lowest-order process
require subtractions in the collinear region that make this differential
cross section a distribution instead of a function of the variable $v$.
It is therefore more reasonable to consider an observable $A(v)$ and
to integrate over $v$
\beq
\langle A \rangle = \int_0^1 \dd v \, \frac{\dd \sigma}{\dd v} \, A(v).
\eeq
This way one defines the expectation value
\beq
{\cal A}\,=\,\frac{\dd \langle A \rangle}{\dd\XB\,\dd y\,\dd z}.
\eeq
Explicitly, it is given by
\beqn
{\cal A}&=&\sum_j \int \frac{\dd u}{u} \sum_N \sum_{\underline{i}}
\int \dPS^{(N)}(\underline p)\,
\frac{\alpha^2}{2\SH\XB}\,\frac{1}{e^2 (2\pi)^d}\nonu
&\cdot&\!\!\!\!\left[Y^M (-g^{\mu\nu})
            +Y^L\,\frac{4\XB^2}{Q^2}\,\PP^\mu\PP^\nu
       \right]\,H_{\mu\nu}\nonu
&\cdot&\!\!\!\!\Bigg\{
 M_{j,h/\AAA}\left(\frac{\XB}{u},\frac{\EH}{\EE}\right) (1-\XB) A(1)
+f_{j/\AAA}\left(\frac{\XB}{u}\right)\,
\sum_{\alpha=1}^N\,D_{\hh/i_\alpha}\,\left(\frac{\EH}{E_\alpha}\right)
\,\frac{\EE}{E_\alpha} (1-\XB) A(v_\alpha)
       \Bigg\}.
\eeqn
The notation in this formula is the following:
$N$ is the number of final-state partons in the QCD subprocess,
$u=\XB/\xi$, $\underline{i}$ is a multi-index specifying the
parton configuration, so $\underline{i}=(i_1,i_2,\ldots,i_N)$, 
where $i_\alpha \in \{u,\overline{u},d,\overline{d},\ldots,g\}$,
$\underline{p}=(p_1,p_2,\ldots,p_N)$ is the set of 
momenta of the outgoing partons,
$E_\alpha$ is the energy of the $\alpha^{\mbox{\scriptsize th}}$ parton
in the $(\vec{\PP}+\vec{q}=\vec{0})$ frame, and 
$v_\alpha=(1-\cos\vartheta_\alpha)/2$, where $\vartheta_\alpha$ is the polar
angle of the $\alpha^{\mbox{\scriptsize th}}$ parton in the same frame;
$M$, $f$ and $D$ are the fracture functions, the parton densities
and the fragmentation functions, respectively.

\noindent
It is convenient to define the following constants and projection operators:
\beqn
c_i&=&\frac{\alpha^2}{2\SH \XB}\cdot 2\pi \cdot 4(1-\epsilon) \, Q_i^2,\\
\Pii_M^{\mu\nu}&=&\left(-g^{\mu\nu}\right),\quad
\Pii_L^{\mu\nu}=\frac{4\XB^2}{Q^2}\PP^\mu\PP^\nu.
\eeqn
Here, $eQ_i$ is the electric charge of the quark of flavour $i$.

\noindent
The leading order given by the processes of the parton-model depicted
in Figs.~2~(a) and 2~(b) is
\beqn
{\cal A}_{\mbox{\scriptsize LO}}&=&Y^M\,\sum_{i=q,\overline{q}}\,c_i\nonu
&\cdot&\Bigg\{
   \int_{\frac{\XB}{1-(1-\XB)z}}^1\frac{\dd u}{u}M_{i,\hh/\AAA}
\left(\frac{\XB}{u},(1-\XB)z\right)\,\delta(1-u)\,(1-\XB)\,A(1)\nonu
& &+
\int_{\XB}^1\frac{\dd u}{u}\,
\int\frac{\dd \rho}{\rho}\,
f_{i/\AAA}
\left(\frac{\XB}{u}\right)
\,D_{\hh/i}\left(\frac{z}{\rho}\right)\,\delta(1-u)\,\delta(1-\rho)\,A(0)
\Bigg\}
\eeqn
To this order, there are no longitudinal contributions. They arise
in the QCD-improved parton-model in \porder{\alpha_s}, which is 
considered in the next section.

\section{The Cross Section to \porder{\alpha_s}}
\label{sect3}
\noindent
This section gives the details of the calculation of the
\porder{\alpha_s} corrections
to the parton-model processes. The virtual 1-loop corrections 
to the leading-order QCD subprocess are shown in Fig.~3, and the
real corrections in Figs.~4~(a),(b).

\noindent
The overall effect of the virtual corrections is to multiply the
leading-order cross section by \cite{16}
\beq
1\,+\,\frac{\alpha_s}{2\pi}
\left(\frac{4\pi\mu^2}{Q^2}\right)^\epsilon
\frac{\Gamma(1-\epsilon)}{\Gamma(1-2\epsilon)}C_F
\left(-2\frac{1}{\epsilon^2}-3\frac{1}{\epsilon}-8-\frac{\pi^2}{3}
\right),
\eeq
so
\beqn
{\cal A}_{\mbox{\scriptsize LO+virt.}}
&=&Y^M\,\sum_{i=q,\overline{q}}\,c_i\nonu
&\cdot&\Bigg\{
   \int_{\frac{\XB}{1-(1-\XB)z}}^1\frac{\dd u}{u}M_{i,\hh/\AAA}
\left(\frac{\XB}{u},(1-\XB)z\right)\,\delta(1-u)\,(1-\XB)\,A(1)\nonu
& &+
\int_{\XB}^1\frac{\dd u}{u}\,
\int\frac{\dd \rho}{\rho}\,
f_{i/\AAA}
\left(\frac{\XB}{u}\right)
\,D_{\hh/i}\left(\frac{z}{\rho}\right)\,\delta(1-u)\,\delta(1-\rho)\,A(0)
\Bigg\}\nonu
& &\cdot\Bigg\{
1\,+\,\frac{\alpha_s}{2\pi}
\left(\frac{4\pi\mu^2}{Q^2}\right)^\epsilon
\frac{\Gamma(1-\epsilon)}{\Gamma(1-2\epsilon)}C_F
\left(-2\frac{1}{\epsilon^2}-3\frac{1}{\epsilon}-8-\frac{\pi^2}{3}
\right)\Bigg\},
\eeqn
where $\mu$ is the renormalization scale, 
$\alpha_s=\alpha_s(\mu^2)$, and $C_F=4/3$ is one of the Casimir invariants
of the colour gauge group SU(3).

\noindent
The double and single poles in $\epsilon$ represent an infrared divergence,
which is cancelled by a contribution similar
to the real corrections but of opposite sign.

\noindent
The projections of the hadron tensor for the real corrections are
(see, for example, \cite{17})
\beqn
&&\frac{1}{e^2(2\pi)^{2d}}\Pii_M^{\mu\nu}H_{\mu\nu}(\gamma^*q\rightarrow qg)\nonu
&&\quad=\,8\pi\, \frac{\alpha_s}{2\pi}\,\mu^{2\epsilon}
\,2\pi\,C_F\,Q_q^2\,4(1-\epsilon)\,
\left[
  (1-\epsilon)\left(
       \frac{s_{ig}}{s_{qg}}
      +\frac{s_{qg}}{s_{ig}}
              \right)
 +\frac{2Q^2s_{iq}}{s_{ig}s_{qg}}
 +2\epsilon
\right]\\
&&\frac{1}{e^2(2\pi)^{2d}}\Pii_L^{\mu\nu}H_{\mu\nu}(\gamma^*q\rightarrow qg)\nonu
&&\quad=\,8\pi\, \frac{\alpha_s}{2\pi}\,\mu^{2\epsilon}
\,2\pi\,C_F\,Q_q^2\,4(1-\epsilon)\,
\left[
4\,\frac{u^2}{Q^2}\,\frac{1}{2}s_{iq}
\right]\\
&&\frac{1}{e^2(2\pi)^{2d}}\Pii_M^{\mu\nu}H_{\mu\nu}(\gamma^*g\rightarrow q\qb)
\nonu
&&\quad=\,8\pi\, \frac{\alpha_s}{2\pi}\,\mu^{2\epsilon}
\,2\pi\,\frac{1}{2}\,Q_q^2\,4(1-\epsilon)\,
\left[
       \frac{s_{iq}}{s_{i\qb}}
      +\frac{s_{i\qb}}{s_{iq}}
 -\frac{1}{1-\epsilon}\frac{2Q^2s_{q\qb}}{s_{iq}s_{i\qb}}
 -2\frac{\epsilon}{1-\epsilon}
\right]\\
&&\frac{1}{e^2(2\pi)^{2d}}\Pii_L^{\mu\nu}H_{\mu\nu}(\gamma^*g\rightarrow q\qb)
\nonu
&&\quad=\,8\pi\, \frac{\alpha_s}{2\pi}\,\mu^{2\epsilon}
\,2\pi\,\frac{1}{2}\,Q_q^2\,4(1-\epsilon)\,
\left[
4\,\frac{u^2}{Q^2}\,s_{q\qb}
\right].
\eeqn
The invariants are
$s_{AB}=2p_A p_B$, 
and $p_i$, $p_q$, $p_{\qb}$, $p_g$ are the momenta of
the incoming parton (quark or gluon), an outgoing quark, an outgoing
antiquark and an outgoing gluon, respectively. The formulas already contain
the appropriate factors for the average over the colour degrees of freedom
of the incoming partons. An additional factor of $1/(1-\epsilon)$
has been provided for the terms with an incoming gluon,
because gluons have $2(1-\epsilon)$ helicity states in
$4-2\epsilon$ space-time dimensions compared to only two in 
the case of quarks.
\noindent
In order to perform the phase-space integrations, suitable parametrizations
\beq
\dPS^{(2)}\,=\,(2\pi)^d\prod_{i=1}^{2}\,\left(
\frac{\dd p_i\,\delta(p_i^2)}{(2\pi)^{d-1}}
\right)
\eeq
of the two-particle phase-space are needed. 
The integration variable has been chosen such that it is the
one that is actually used in the factorization of the collinear singularities.
The integration over irrelevant angles can be performed. 
The results for three phase-space parametrizations A, B and C
in terms of the variables $\rho$, $w$, $u'$
are written out in Appendix \ref{appps}.

\noindent
By means of these phase-space parametrizations, the integrations can
be performed and the infrared and collinear
singularities can be factorized.
The details of this straightforward but cumbersome calculation
are not given here. 
The results are presented in a form that explicitly shows
the cancellation of the collinear singularities by means of a renormalization
of the phenomenological distribution functions.
The contributions are denoted by
$B_1^M$, $B_2^M$, $C^M$,
$B_1^L$, $B_2^L$, $C^L$.
Here the indices in $B^A_\alpha$ and $C^A$ stand for
\begin{itemize}
\item $\alpha$: phase-space region for the integration variable $u$,\\
$\alpha=1$: 
$u\in\left[\XB,\frac{\XB}{\XB+(1-\XB)z}\right]$,\\
$\alpha=2$:
$u\in\left[\frac{\XB}{\XB+(1-\XB)z},1\right]$;
\item
$A$: polarization of the virtual photon,\\
$A=M$: metric contribution,\\
$A=L$: longitudinal contribution. 
\end{itemize}
The observed hadron in the terms $B^A_\alpha$ 
originates from one of the outgoing partons via a 
fragmentation function, 
the one in the terms $C^A$ 
comes from the incoming hadron via a fracture function.
The $B^A_\alpha$ and $C^A$ terms receive contributions from the graphs from
Figs.~4~(a) and (b), respectively.
Throughout the calculation the $\overline{\mbox{MS}}$ factorization
scheme both for the parton densities and fracture functions and for
the fragmentation functions is used. The choice of 
the factorization scheme defines
the finite parts unambiguously. One has to choose two factorization scales, 
one ($M_f^2$) for the renormalized distribution functions $f$ and $M$ for partons in
the incoming nucleon, and one ($M_D^2$) for the fragmentation functions $D$.

\noindent
The explicit results for the cross sections are collected in 
Appendix \ref{appsummary}.
One sees immediately that the IR singularities proportional to
$2/\epsilon^2+3/\epsilon$ cancel among virtual and real corrections.
The cancellation of the remaining collinear singularities  is discussed 
in the next section, and the finite terms $\Phi_X^A$ 
from appendix \ref{appsummary} are discussed in 
Section~\ref{finite}.

\section{Renormalized Distribution Functions}
\label{sect4}
\label{renden}
The renormalization of the bare parton densities $f_{i/\AAA}$ and 
bare fragmentation functions $D_{h/i}$ is done in the 
$\overline{\mbox{MS}}$ factorization scheme in the
standard way \cite{16,7}, as required by the total cross section
in lepton--nucleon scattering and the one-particle inclusive cross section
in $e^+e^-$ scattering. The expressions for the bare densities in terms 
of the renormalized ones for the factorization scales
$M_f^2$ and $M_D^2$ are
\beqn
f_{i/\AAA}(\xi)&=&\int_\xi^1\frac{\dd u}{u}\,
\left[\delta_{ij}\,\delta(1-u)
+\frac{1}{\epsilon}\frac{\alpha_s}{2\pi}
\frac{\Gamma(1-\epsilon)}{\Gamma(1-2\epsilon)}
\left(\frac{4\pi\mu^2}{M_f^2}\right)^\epsilon
P_{i\leftarrow j}(u)\right]
\,f^r_{j/\AAA}\left(\frac{\xi}{u},M_f^2\right)
\\
D_{\hh/i}(\xi)&=&\int_\xi^1\frac{\dd u}{u}\,
\left[\delta_{ij}\,\delta(1-u)
+\frac{1}{\epsilon}\frac{\alpha_s}{2\pi}
\frac{\Gamma(1-\epsilon)}{\Gamma(1-2\epsilon)}
\left(\frac{4\pi\mu^2}{M_D^2}\right)^\epsilon
P_{j\leftarrow i}(u)\right]
\,D^r_{\hh/j}\left(\frac{\xi}{u},M_D^2\right).
\eeqn
The expression for $M_{i,\hh/\AAA}(\xi,\zeta)$
in terms of the renormalized quantity 
$M^r_{i,\hh/\AAA}(\xi,\zeta,M_f^2)$
can be obtained by the requirement that all 
collinear divergences that are not already absorbed into
$f^r$ and $D^r$ are absorbed into $M^r$.
By an appropriate definition of the finite parts, the
bare fracture function in terms of the renormalized fracture function
in the 
$\overline{\mbox{MS}}$ factorization scheme  
is
\beqn
\!\!\!\!\!
M_{i,\hh/\AAA}(\xi,\zeta)&=&\int_\frac{\xi}{1-\zeta}^1\frac{\dd u}{u}\,
\left[\delta_{ij}\,\delta(1-u)
+\frac{1}{\epsilon}\frac{\alpha_s}{2\pi}
\frac{\Gamma(1-\epsilon)}{\Gamma(1-2\epsilon)}
\left(\frac{4\pi\mu^2}{M_f^2}\right)^\epsilon
P_{i\leftarrow j}(u)\right]
\,M^r_{j,\hh/\AAA}\left(\frac{\xi}{u},\zeta,M_f^2\right)\nonu
&&\!\!\!\!\!\!\!\!\!\!\!\!\!\!\!\!\!\!\!\!\!\!\!\!\!\!\!\!\!\!\!\!\!\!+
\int_\xi^\frac{\xi}{\xi+\zeta}\frac{\dd u}{u}\,
\frac{1}{1-u}
\,\frac{u}{\XB}\,
\frac{1}{\epsilon}\frac{\alpha_s}{2\pi}
\frac{\Gamma(1-\epsilon)}{\Gamma(1-2\epsilon)}
\left(\frac{4\pi\mu^2}{M_f^2}\right)^\epsilon
\hat{P}_{ki\leftarrow j}(u)
\,f_{j/\AAA}\left(\frac{\xi}{u}\right)
\,D_{\hh/k}\left(\frac{\zeta u}{\xi(1-u)}\right).
\eeqn
We have implicitly assumed that repeated indices
are summed over.
These expressions are valid to \porder{\alpha_s}.
The formula for $M$ is inhomogeneous. The homogeneous
contribution is just the standard Altarelli-Parisi evolution
caused by the emission of collinear partons from the parton 
emanating from a fracture function. The inhomogeneous term arises
because the observed hadron may originate from a fragmentation function
of an outgoing parton collinear to the incoming one.

\noindent
By requiring that the bare fracture functions do not depend on the
factorization scale and consequently
\beq
\frac{\dd}{\dd \ln M_f^2}\,M_{i,\hh/\AAA}(\xi,\zeta)=0,
\eeq
the evolution equation for
$M^r_{i,\hh/\AAA}(\xi,\zeta,M_f^2)$ can be derived. 
It coincides with that
in \cite{14}.

\noindent
Rewriting the bare distribution functions in terms of the
renormalized ones in all cross-section formulas, and keeping
only terms up to \porder{\alpha_s}, one can see that
all collinear singularities cancel. 
The result is a finite cross section to
\porder{\alpha_s}, which is discussed in the next section.
 
\section{Finite Contributions}
\label{finite}
After the cancellation of the infrared and collinear singularities
the resulting cross section is finite. Because of the
subtractions, it is of the form of a convolution of a distribution
with phenomenological distribution functions.
The ``+''~prescriptions used are defined by \cite{18}:
\beq
\int \dd x \, D_{+x[\underline{a},b]}(x)\,\varphi(x)
=\int_a^b\,\dd x \,D(x)\,(\varphi(x)-\varphi(a)).
\eeq
The variable
and the range of the integration
are indicated as a subscript; furthermore, the subtraction point
is underlined. These distributions are used in the expressions
of the $\Phi_X^{L/M}$, which are presented in Appendix \ref{appexplicit}.
The expectation value of the observable $A$ is, differential in
$\XB$, $y$ and $z$:
\beq
\frac{\dd\,\langle A\rangle}
{\dd\XB\,\dd y\,\dd z }=
 {\cal A}_{\mbox{\scriptsize LO+virt.}}^f
+{\cal A}_{B_1^M}^f
+{\cal A}_{B_2^M}^f
+{\cal A}_{C^M}^f
+{\cal A}_{B_1^L}^f
+{\cal A}_{B_2^L}^f
+{\cal A}_{C^L}^f,
\eeq
with
\setlength{\conserve}{\mathindent}
\setlength{\mathindent}{\wideformula}
\beqn
{\cal A}^f_{\mbox{\scriptsize LO+virt.}}
&=&Y^M\,\sum_{i=q,\overline{q}}\,c_i\nonu
&\cdot&\Bigg\{
   \int_{\frac{\XB}{1-(1-\XB)z}}^1\frac{\dd u}{u}M^r_{i,\hh/\AAA}
\left(\frac{\XB}{u},(1-\XB)z,M_f^2\right)\,\delta(1-u)\,(1-\XB)\,A(1)\nonu
& &+
\int_{\XB}^1\frac{\dd u}{u}\,
\int\frac{\dd \rho}{\rho}\,
f^r_{i/\AAA}
\left(\frac{\XB}{u},M_f^2\right)
\,D^r_{\hh/\AAA}\left(\frac{z}{\rho},M_D^2\right)\,\delta(1-u)\,
\delta(1-\rho)\,A(0)
\Bigg\}\nonu
& &\cdot\Bigg\{
1\,+\,\frac{\alpha_s}{2\pi}
\left(-8-\frac{\pi^2}{3}
\right)\Bigg\},
\eeqn
\setlength{\mathindent}{\conserve}
\setlength{\conserve}{\mathindent}
\setlength{\mathindent}{\wideformula}
\beqn
{\cal A}_{B^M_1}^f&=&Y^M\,\sum_{i=q,\qb}\,c_i\,\frac{\alpha_s}{2\pi}\,
\int_{\XB}^{\frac{\XB}{\XB+(1-\XB)z}}\,\frac{\dd u}{u}\,
\int_{a(u)}^1\,\frac{\dd \rho}{\rho}\,A(v(\rho,u))\nonu
&&\cdot\Bigg[
f^r_{i/\AAA}\left(\frac{\XB}{u},M_f^2\right)\,
D^r_{\hh/i}\left(\frac{z}{\rho},M_D^2\right)\,
\left\{
-\ln\frac{M_f^2}{Q^2}\,P_{q\leftarrow q}(u)\,\delta(1-\rho)
+C_F\,\Phi_{1qq}^M
\right\}\nonu
&&\;\;+
f^r_{i/\AAA}\left(\frac{\XB}{u},M_f^2\right)\,
D^r_{\hh/g}\left(\frac{z}{\rho},M_D^2\right)\,
\left\{
-\ln\frac{M_f^2}{Q^2}\,\hat{P}_{gq\leftarrow q}(u)\,\delta(\rho-a(u))
+C_F\,\Phi_{1qg}^M
\right\}\nonu
&&\;\;+
f^r_{g/\AAA}\left(\frac{\XB}{u},M_f^2\right)\,
D^r_{\hh/i}\left(\frac{z}{\rho},M_D^2\right)\nonu
&&\;\;
\cdot\left\{
-\ln\frac{M_f^2}{Q^2}\,\hat{P}_{\qb q\leftarrow g}(u)\,\delta(\rho-a(u))
-\ln\frac{M_f^2}{Q^2}\,P_{q\leftarrow g}(u)\,\delta(1-\rho)
+\frac{1}{2}\,\Phi_{1gq}^M
\right\}\Bigg]
\eeqn
\setlength{\mathindent}{\conserve}
\setlength{\conserve}{\mathindent}
\setlength{\mathindent}{\wideformula}
\beqn
{\cal A}^f_{B^M_2}&=&Y^M\,\sum_{i=q,\qb}\,c_i\,\frac{\alpha_s}{2\pi}\,
\int_{\XB}^{\frac{\XB+(1-\XB)z}{1}}\,\frac{\dd u}{u}\,
\int_{z}^1\,\frac{\dd \rho}{\rho}\,A(v(\rho,u))\nonu
&&\cdot\Bigg[
f^r_{i/\AAA}\left(\frac{\XB}{u},M_f^2\right)\,
D^r_{\hh/i}\left(\frac{z}{\rho},M_D^2\right)\nonu
&&\;\;\cdot\left\{
-\ln\frac{M_f^2}{Q^2}\,P_{q\leftarrow q}(u)\,\delta(1-\rho)
-\ln\frac{M_D^2}{Q^2}\,P_{q\leftarrow q}(\rho)\,\delta(1-u)
+C_F\,\Phi_{2qq}^M
\right\}\nonu
&&\;+
f^r_{i/\AAA}\left(\frac{\XB}{u},M_f^2\right)\,
D^r_{\hh/g}\left(\frac{z}{\rho},M_D^2\right)\,
\left\{
-\ln\frac{M_D^2}{Q^2}\,P_{g\leftarrow q}(\rho)\,\delta(1-u)
+C_F\,\Phi_{2qg}^M
\right\}\nonu
&&\;+
f^r_{g/\AAA}\left(\frac{\XB}{u},M_f^2\right)\,
D^r_{\hh/i}\left(\frac{z}{\rho},M_D^2\right)\,
\left\{
-\ln\frac{M_f^2}{Q^2}\,P_{q\leftarrow g}(u)\,\delta(1-\rho)
+\frac{1}{2}\,\Phi_{2gq}^M
\right\}
\Bigg]
\eeqn
\setlength{\mathindent}{\conserve}
\setlength{\conserve}{\mathindent}
\setlength{\mathindent}{\wideformula}
\beqn
{\cal A}^f_{C^M}&=&Y^M\,\sum_{i=q,\qb}\,c_i\,\frac{\alpha_s}{2\pi}\,
\int_{\frac{\XB}{1-(1-\XB)z}}^1\,\frac{\dd u}{u}\,A(1)\nonu
&&\cdot\Bigg[
M^r_{i,h/\AAA}\left(\frac{\XB}{u},(1-\XB)z,M_f^2\right)\,
\bigg\{
-\ln\frac{M_f^2}{Q^2}\,P_{q\leftarrow q}(u)\,(1-\XB)
+C_F\,\Phi_{q}^M
\bigg\}\nonu
&&\;+
M^r_{g,h/\AAA}\left(\frac{\XB}{u},(1-\XB)z,M_f^2\right)\,
\bigg\{
-\ln\frac{M_f^2}{Q^2}\,P_{q\leftarrow g}(u)\,(1-\XB)
+\frac{1}{2}\,\Phi_{g}^M
\bigg\}
\Bigg]
\eeqn
\setlength{\mathindent}{\conserve}
\setlength{\conserve}{\mathindent}
\setlength{\mathindent}{\wideformula}
\beqn
{\cal A}^f_{B^L_1}&=&Y^L\,\sum_{i=q,\qb}\,c_i\,\frac{\alpha_s}{2\pi}\,
\int_{\XB}^{\frac{\XB}{\XB+(1-\XB)z}}\,\frac{\dd u}{u}\,
\int_{a(u)}^1\,\frac{\dd \rho}{\rho}\,A(v(\rho,u))\nonu
&&\cdot\Bigg[
f^r_{i/\AAA}\left(\frac{\XB}{u},M_f^2\right)\,
D^r_{\hh/i}\left(\frac{z}{\rho},M_D^2\right)\,
C_F\,\Phi_{1qq}^L\nonu
&&+
f^r_{i/\AAA}\left(\frac{\XB}{u},M_f^2\right)\,
D^r_{\hh/g}\left(\frac{z}{\rho},M_D^2\right)\,
C_F\,\Phi_{1qg}^L\nonu
&&+
f^r_{g/\AAA}\left(\frac{\XB}{u},M_f^2\right)\,
D^r_{\hh/i}\left(\frac{z}{\rho},M_D^2\right)\,
\frac{1}{2}\,\Phi_{1gq}^L
\Bigg]
\eeqn
\setlength{\mathindent}{\conserve}
%\newpage
\setlength{\conserve}{\mathindent}
\setlength{\mathindent}{\wideformula}
\beqn
{\cal A}^f_{B^L_2[A}&=&Y^L\,\sum_{i=q,\qb}\,c_i\,\frac{\alpha_s}{2\pi}\,
\int_{\frac{\XB}{\XB+(1-\XB)z}}^1\,\frac{\dd u}{u}\,
\int_{z}^1\,\frac{\dd \rho}{\rho}\,A(v(\rho,u))\nonu
&&\cdot\Bigg[
f^r_{i/\AAA}\left(\frac{\XB}{u},M_f^2\right)\,
D^r_{\hh/i}\left(\frac{z}{\rho},M_D^2\right)\,
C_F\,\Phi_{2qq}^L\nonu
&&+
f^r_{i/\AAA}\left(\frac{\XB}{u},M_f^2\right)\,
D^r_{\hh/g}\left(\frac{z}{\rho},M_D^2\right)\,
C_F\,\Phi_{2qg}^L\nonu
&&+
f^r_{g/\AAA}\left(\frac{\XB}{u},M_f^2\right)\,
D^r_{\hh/i}\left(\frac{z}{\rho},M_D^2\right)\,
\frac{1}{2}\,\Phi_{2gq}^L
\Bigg]
\eeqn
\setlength{\mathindent}{\conserve}
\setlength{\conserve}{\mathindent}
\setlength{\mathindent}{\wideformula}
\beqn
{\cal A}^f_{C^L}&=&Y^L\,\sum_{i=q,\qb}\,c_i\,\frac{\alpha_s}{2\pi}\,
\int_{\frac{\XB}{1-(1-\XB)z}}^1\,\frac{\dd u}{u}\,A(1)\nonu
&&\cdot\Bigg[
M^r_{i,h/\AAA}\left(\frac{\XB}{u},(1-\XB)z,M_f^2\right)\,
C_F\,\Phi_{q}^L\nonu
&&+
M^r_{g,h/\AAA}\left(\frac{\XB}{u},(1-\XB)z,M_f^2\right)\,
\frac{1}{2}\,\Phi_{g}^L
\Bigg]
\eeqn
\setlength{\mathindent}{\conserve}
For some of the contributions the integration region is split into
two parts, according to finite terms that arise from singular regions.
The form given here should be suitable for numerical evaluation.

\section{Summary and Conclusions}
In this paper the one-particle inclusive cross section 
in deeply inelastic lepton--nucleon scattering has 
been expressed in terms of parton densities, fragmentation functions
and fracture functions.
A cut in the transverse momentum of the observed particle is not required.
Compared to the case
of a hadron that is well separated from the target remnant,
additional collinear singularities are encountered
because of the enlarged phase-space
available to the parent parton of the observed particle.
They stem from the phase-space region where the parent parton of the observed 
particle is collinear to the incoming parton.
These singularities can, to \porder{\alpha_s}, be consistently
absorbed into the fracture functions.
A check of the calculation is provided by the determination of
the evolution equation of the fracture functions; it coincides
with the one given in \cite{14}.

\noindent
Fracture functions allow for a coherent and unified description of
one-particle inclusive production processes in the case of
incoming hadrons, without having
to assume specific models for the target fragmentation \cite{19}.
As in the case of parton densities, 
the ignorance of the nucleon wave function is parametrized
in a phenomenological distribution function, 
which has to measured.
For target fragmentation, 
fracture functions have the advantage over
phenomenological models
that the correct scale dependence
is built in, because of their evolution equation.
It is expected that fracture functions are as 
universal as parton densities, i.e. that they can be determined in a 
specific process and then used for predictions
in other processes.

\noindent
The results of this paper should be consistent to \porder{\alpha_s}
if the fracture functions are extracted at a scale $M_0^2$
and then used for predictions at a scale $M^2$, with 
$\ln\left(M^2/M_0^2\right)$ not too large.
However, in order to evolve the fracture functions to scales $M^2$ 
very different from $M_0^2$, the evolution kernel
(the analogon of the anomalous dimensions in the case of parton densities
and fragmentation functions) should be known on the two-loop level.
This effort would be necessary to have genuine \porder{\alpha_s}
predictions in this case.

\noindent
An interesting question would be to find out whether factorization
of QCD subprocesses is always possible that collinear
singularities can consistently be absorbed order by order
into renormalized distribution functions, in the case that 
there is no transverse momentum cut for the observed particle.
This would certainly be necessary to make the cross section
well-defined.
Another question is whether there is a connection with the formalism
of Mueller \cite{20,21,22} for one-particle inclusive processes.
We hope to return to these questions in the future.

\noindent
A more practical issue is the application of the concepts
of \cite{14} and the results of this paper to a real
experimental situation, e.g. to deeply inelastic
electron-proton scattering at HERA. 
An interesting application would be the ``heavy-quark inclusive'' cross
section, i.e. the cross section for the production of a heavy quark
plus anything. There is no obvious reason why the methods used
in this paper should not be applicable to the production of
heavy quarks instead of hadrons. In fact, explicit results for
the fragmentation function of partons into heavy quarks are available
\cite{23}, and similar results should be possible for
fracture functions. Work along these lines is in progress.

\vspace{1cm}
\noindent
{\Large \bf Acknowledgements}
\begin{sloppypar}
I gratefully acknowledge disscussions with A.~Ali, G.~Altarelli,
P.~Nason, T.~Sj\"{o}strand and G.~Veneziano.
Moreover I 
wish to thank P.~Nason and G.~Veneziano for a critical reading
of the manuscript.
\end{sloppypar}

\newpage
\begin{appendix}
\section{Phase-Space Parametrizations}
\label{appps}
\noindent 
In this appendix three different parametrizations of the two-particle
phase-space are given.

\noindent
A) The integration variable $\rho$
is the energy $E_1$ of one of the partons
in the $(\vec{\PP}+\vec{q}=\vec{0})$ frame, scaled by the proton 
momentum,
\beq
\rho\,=\,\frac{E_1}{\EE(1-\XB)}.
\eeq
This parametrization is used for those contributions in which  a 
collinear singularity has to be absorbed into a fragmentation function $D$;
it is
\beqn
\dPS^{(2)}\,&=&\,\frac{1}{8\pi}\,
\frac{(4\pi)^\epsilon}{\Gamma(1-\epsilon)}\,
\left(Q^2\right)^{-\epsilon}\,
\frac{u(1-\XB)}{u-\XB}\,\nonu
&\cdot&(1-\XB)^{-2\epsilon}\,u^{-\epsilon}\,
(1-u)^{-\epsilon}\,(u-\XB)^{2\epsilon}\,
(\rho-a(u))^{-\epsilon}\,(1-\rho)^{-\epsilon}\,\dd \rho.
\eeqn
Here 
\beq
a(u)=\frac{\XB}{1-\XB}\frac{1-u}{u}.
\eeq
The energies and invariants are given by
\beqn
E_1&=&\EE(1-\XB)\rho\\
E_2&=&\EE(1-\XB)(1-\rho+a(u))\\
s_{12}&=&Q^2\frac{1-u}{u}\\
s_{i1}&=&Q^2\frac{1-\XB}{u-\XB}(\rho-a(u))\\
s_{i2}&=&Q^2\frac{1-\XB}{u-\XB}(1-\rho).
\eeqn
The angular variable $v_1=(1-\cos\vartheta_1)/2$ is
\beq
v_1=v(\rho,u)=\frac{\XB(1-u)}{u-\XB} \frac{1-\rho}{\rho}.
\eeq
The range of integration is restricted to
$\rho\in[a(u),1]$;
however, in order to ensure that the energy of the parent parton 
of the observed hadron is larger than that of the hadron,
the additional condition $\rho \geq z$ must be satisfied.

\noindent 
B) The integration variable $w$
is related to an angular variable in the CM system of the
virtual photon and the incoming parton. 
Its relation to $\rho$ is
\beq
w=\frac{1-\rho}{1-a(u)}.
\eeq
This parametrization is used for the contributions that involve a
fracture function;
it is given by 
\beqn
\dPS^{(2)}\,&=&\,\frac{1}{8\pi}\,
\frac{(4\pi)^\epsilon}{\Gamma(1-\epsilon)}\,
\left(Q^2\right)^{-\epsilon}\,
(1-u)^{-\epsilon}\,u^\epsilon\,
(w(1-w))^{-\epsilon}
\dd w.
\eeqn
The energies and invariants are
\beqn
E_1&=&\EE(1-\XB)(1-(1-a(u))w)\\
E_2&=&\EE(1-\XB)(a(u)+(1-a(u))w)\\
s_{12}&=&Q^2\frac{1-u}{u}\\
s_{i1}&=&Q^2\frac{1}{u}(1-w)\\
s_{i2}&=&Q^2\frac{1}{u}w.
\eeqn
The angular variable $v_1$ is
\beq
v_1=v(w,u)=\frac{a(u)w}{1-(1-a(u))w}.
\eeq
The range of integration is restricted to
$w\in[0,1]$.

\noindent 
C) This parametrization is convenient for the contributions in which 
the observed hadron originates from a parton that is collinear 
to the incoming parton.
A variable $u'$ is introduced by
\beq
u'=1-\frac{1-\XB}{\XB}\,\rho\,u.
\eeq
The parametrization is given by 
\beqn
\dPS^{(2)}\,&=&\,\frac{1}{8\pi}\,
\frac{(4\pi)^\epsilon}{\Gamma(1-\epsilon)}\,
\left(Q^2\right)^{-\epsilon}\,
\frac{\XB}{u-\XB}\,\nonu
&\cdot&(u-u')^{-\epsilon}\,
(1-u)^{-\epsilon}\,
\left(
 1-\frac{\XB}{1-\XB}\frac{1-u'}{u}
\right)^{-\epsilon}
(u-\XB)^{2\epsilon}\,
\XB^{-\epsilon}\,(1-\XB)^{-\epsilon}\,\dd u'.
\eeqn
The energies and invariants are
\beqn
E_1&=&\EE\,\XB \frac{1-u'}{u}\\
E_2&=&\EE(1-\XB)\left(
 1-\frac{\XB}{1-\XB}\frac{u-u'}{u}
\right)\\
s_{12}&=&Q^2\frac{1-u}{u}\\
s_{i1}&=&Q^2\frac{\XB}{u-\XB}\frac{u-u'}{u}\\
s_{i2}&=&Q^2\frac{1-\XB}{u-\XB}\left(
 1-\frac{\XB}{1-\XB}\frac{1-u'}{u}
\right).
\eeqn
The angular variable $v_1$ is
\beq
v_1=v(u',u)=\frac{1-u}{u-\XB} \frac{(1-\XB)u-\XB(1-u')}{1-u'}.
\eeq
The range of integration is restricted to
$u'\in[1-\frac{1-\XB}{\XB}u,u]$.
It has to be further restricted by
\beq
u'\,\leq\,1-\frac{1-\XB}{\XB}\,z\,u
\eeq
in order to avoid that the outgoing hadron
has a larger energy than its parent parton.

\newpage
\section{Summary of Cross-Section Formulas}
\label{appsummary}
In this appendix the expressions for the 
cross section including the divergent parts are written out. They are
\setlength{\conserve}{\mathindent}
\setlength{\mathindent}{\wideformula}
\beqn
{\cal A}_{B^M_1}&=&Y^M\,\sum_{i=q,\qb}\,c_i\,\frac{\alpha_s}{2\pi}\nonu
&\cdot&\!\!\!\!\Bigg\{
\int_{\XB}^{\frac{\XB}{\XB+(1-\XB)z}}\,\frac{\dd u}{u}\,
\int_{a(u)}^1\,\frac{\dd \rho}{\rho}\,A(v(\rho,u))\nonu
&&\cdot\Bigg[
f_{i/\AAA}\left(\frac{\XB}{u}\right)\,
D_{\hh/i}\left(\frac{z}{\rho}\right)\nonu
&&\;\;\cdot\bigg\{
-\frac{1}{\epsilon}\frac{\Gamma(1-\epsilon)}{\Gamma(1-2\epsilon)}
\left(\frac{4\pi\mu^2}{M_f^2}\right)^\epsilon\,
P_{q\leftarrow q}(u)\,\delta(1-\rho)\nonu
&&\quad\quad
-\ln\frac{M_f^2}{Q^2}\,P_{q\leftarrow q}(u)\,\delta(1-\rho)
+C_F\,\Phi_{1qq}^M
\bigg\}\nonu
&&\;\;+
f_{i/\AAA}\left(\frac{\XB}{u}\right)\,
D_{\hh/g}\left(\frac{z}{\rho}\right)\nonu
&&\;\;\cdot\bigg\{
-\ln\frac{M_f^2}{Q^2}\,\hat{P}_{gq\leftarrow q}(u)\,\delta(\rho-a(u))
+C_F\,\Phi_{1qg}^M
\bigg\}\nonu
&&\;\;+
f_{g/\AAA}\left(\frac{\XB}{u}\right)\,
D_{\hh/i}\left(\frac{z}{\rho}\right)\nonu
&&\;\;\cdot\bigg\{
-\frac{1}{\epsilon}\frac{\Gamma(1-\epsilon)}{\Gamma(1-2\epsilon)}
\left(\frac{4\pi\mu^2}{M_f^2}\right)^\epsilon\,
P_{q\leftarrow g}(u)\,\delta(1-\rho)\nonu
&&\quad\quad
-\ln\frac{M_f^2}{Q^2}\,\hat{P}_{\qb q\leftarrow g}(u)\,\delta(\rho-a(u))
-\ln\frac{M_f^2}{Q^2}\,P_{q\leftarrow g}(u)\,\delta(1-\rho)
+\frac{1}{2}\,\Phi_{1gq}^M
\bigg\}
\Bigg]\nonu
&&\!\!\!\!+
\int_{\XB}^{\frac{\XB}{\XB+(1-\XB)z}}\,\frac{\dd u}{u}\,
(1-\XB)\,A(1)\nonu
&&\cdot\Bigg[
f_{i/\AAA}\left(\frac{\XB}{u}\right)\,
D_{\hh/g}\left(\frac{(1-\XB)\,z\,u}{\XB(1-u)}\right)
\left(-\frac{1}{\epsilon}\right)\frac{\Gamma(1-\epsilon)}{\Gamma(1-2\epsilon)}
\left(\frac{4\pi\mu^2}{M_f^2}\right)^\epsilon\,
\frac{1}{1-u}\,\frac{u}{\XB}\,\hat{P}_{gq\leftarrow q}(u)\nonu
&&\;\;+
f_{g/\AAA}\left(\frac{\XB}{u}\right)\,
D_{\hh/i}\left(\frac{(1-\XB)\,z\,u}{\XB(1-u)}\right)
\left(-\frac{1}{\epsilon}\right)\frac{\Gamma(1-\epsilon)}{\Gamma(1-2\epsilon)}
\left(\frac{4\pi\mu^2}{M_f^2}\right)^\epsilon\,
\frac{1}{1-u}\,\frac{u}{\XB}\,\hat{P}_{\qb q\leftarrow g}(u)
\Bigg]
\Bigg\}\nonu
&&\!\!\!\!\!\!\!\!\!+\porder{\epsilon}
\eeqn
\setlength{\mathindent}{\conserve}
\setlength{\conserve}{\mathindent}
\setlength{\mathindent}{\wideformula}
\beqn
{\cal A}_{B^M_2}&=&Y^M\,\sum_{i=q,\qb}\,c_i\,\frac{\alpha_s}{2\pi}\,
\int_{\frac{\XB}{\XB+(1-\XB)z}}^1\,\frac{\dd u}{u}\,
\int_{z}^1\,\frac{\dd \rho}{\rho}\,A(v(\rho,u))\nonu
&&\cdot\Bigg[
f_{i/\AAA}\left(\frac{\XB}{u}\right)\,
D_{\hh/i}\left(\frac{z}{\rho}\right)\nonu
&&\;\;\cdot\bigg\{
\frac{\Gamma(1-\epsilon)}{\Gamma(1-2\epsilon)}
\left(\frac{4\pi\mu^2}{Q^2}\right)^\epsilon\,
C_F\,\left(2\frac{1}{\epsilon^2}+3\frac{1}{\epsilon}\right)\,
\delta(1-u)\delta(1-\rho)\nonu
&&\quad\quad
-\frac{1}{\epsilon}\frac{\Gamma(1-\epsilon)}{\Gamma(1-2\epsilon)}
\left[
\left(\frac{4\pi\mu^2}{M_f^2}\right)^\epsilon\,
P_{q\leftarrow q}(u)\,\delta(1-\rho)
+\left(\frac{4\pi\mu^2}{M_D^2}\right)^\epsilon\,
P_{q\leftarrow q}(\rho)\,\delta(1-u)
\right]\nonu
&&\quad\quad
-\ln\frac{M_f^2}{Q^2}\,P_{q\leftarrow q}(u)\,\delta(1-\rho)
-\ln\frac{M_D^2}{Q^2}\,P_{q\leftarrow q}(\rho)\,\delta(1-u)
+C_F\,\Phi_{2qq}^M
\bigg\}\nonu
&&\;+
f_{i/\AAA}\left(\frac{\XB}{u}\right)\,
D_{\hh/g}\left(\frac{z}{\rho}\right)\nonu
&&\;\;\cdot\bigg\{
-\frac{1}{\epsilon}\frac{\Gamma(1-\epsilon)}{\Gamma(1-2\epsilon)}
\left(\frac{4\pi\mu^2}{M_f^2}\right)^\epsilon\,
P_{g\leftarrow q}(\rho)\,\delta(1-u)
\nonu
&&\quad\quad
-\ln\frac{M_D^2}{Q^2}\,P_{g\leftarrow q}(\rho)\,\delta(1-u)
+C_F\,\Phi_{2qg}^M
\bigg\}\nonu
&&\;+
f_{g/\AAA}\left(\frac{\XB}{u}\right)\,
D_{\hh/i}\left(\frac{z}{\rho}\right)\nonu
&&\;\;\cdot\bigg\{
-\frac{1}{\epsilon}\frac{\Gamma(1-\epsilon)}{\Gamma(1-2\epsilon)}
\left(\frac{4\pi\mu^2}{M_f^2}\right)^\epsilon\,
P_{q\leftarrow g}(u)\,\delta(1-\rho)\nonu
&&\quad\quad
-\ln\frac{M_f^2}{Q^2}\,P_{q\leftarrow g}(u)\,\delta(1-\rho)
+\frac{1}{2}\,\Phi_{2gq}^M
\bigg\}
\Bigg]\nonu
&&\!\!\!\!\!\!\!\!\!+\porder{\epsilon}
\eeqn
\setlength{\mathindent}{\conserve}
\setlength{\conserve}{\mathindent}
\setlength{\mathindent}{\wideformula}
\beqn
{\cal A}_{C^M}&=&Y^M\,\sum_{i=q,\qb}\,c_i\,\frac{\alpha_s}{2\pi}\,
\int_{\frac{\XB}{1-(1-\XB)z}}^1\,\frac{\dd u}{u}\,A(1)\nonu
&&\cdot\Bigg[
M_{i,h/\AAA}\left(\frac{\XB}{u},(1-\XB)z\right)\nonu
&&\;\;\cdot\bigg\{
\frac{\Gamma(1-\epsilon)}{\Gamma(1-2\epsilon)}
\left(\frac{4\pi\mu^2}{Q^2}\right)^\epsilon\,
C_F\,\left(2\frac{1}{\epsilon^2}+3\frac{1}{\epsilon}\right)\,
\delta(1-u)\,(1-\XB)\nonu
&&\quad\quad
-\frac{1}{\epsilon}\frac{\Gamma(1-\epsilon)}{\Gamma(1-2\epsilon)}
\left(\frac{4\pi\mu^2}{M_f^2}\right)^\epsilon\,
P_{q\leftarrow q}(u)\,(1-\XB)\nonu
&&\quad\quad
-\ln\frac{M_f^2}{Q^2}\,P_{q\leftarrow q}(u)\,(1-\XB)
+C_F\,\Phi_{q}^M
\bigg\}\nonu
&&\;+
M_{g,h/\AAA}\left(\frac{\XB}{u},(1-\XB)z\right)\nonu
&&\;\;\cdot\bigg\{
-\frac{1}{\epsilon}\frac{\Gamma(1-\epsilon)}{\Gamma(1-2\epsilon)}
\left(\frac{4\pi\mu^2}{M_f^2}\right)^\epsilon\,
P_{q\leftarrow g}(u)\,(1-\XB)\nonu
&&\quad\quad
-\ln\frac{M_f^2}{Q^2}\,P_{q\leftarrow g}(u)\,(1-\XB)
+\frac{1}{2}\,\Phi_{g}^M
\bigg\}
\Bigg]\nonu
&&\!\!\!\!\!\!\!\!\!+\porder{\epsilon}
\eeqn
\setlength{\mathindent}{\conserve}
\setlength{\conserve}{\mathindent}
\setlength{\mathindent}{\wideformula}
\beqn
{\cal A}_{B^L_1}&=&Y^L\,\sum_{i=q,\qb}\,c_i\,\frac{\alpha_s}{2\pi}\,
\int_{\XB}^{\frac{\XB}{\XB+(1-\XB)z}}\,\frac{\dd u}{u}\,
\int_{a(u)}^1\,\frac{\dd \rho}{\rho}\,A(v(\rho,u))\nonu
&&\cdot\Bigg[
f_{i/\AAA}\left(\frac{\XB}{u}\right)\,
D_{\hh/i}\left(\frac{z}{\rho}\right)\,
C_F\,\Phi_{1qq}^L\nonu
&&+
f_{i/\AAA}\left(\frac{\XB}{u}\right)\,
D_{\hh/g}\left(\frac{z}{\rho}\right)\,
C_F\,\Phi_{1qg}^L\nonu
&&+
f_{g/\AAA}\left(\frac{\XB}{u}\right)\,
D_{\hh/i}\left(\frac{z}{\rho}\right)\,
\frac{1}{2}\,\Phi_{1gq}^L
\Bigg]+\porder{\epsilon}
\eeqn
\setlength{\mathindent}{\conserve}
\newpage
\setlength{\conserve}{\mathindent}
\setlength{\mathindent}{\wideformula}
\beqn
{\cal A}_{B^L_2}&=&Y^L\,\sum_{i=q,\qb}\,c_i\,\frac{\alpha_s}{2\pi}\,
\int_{\frac{\XB}{\XB+(1-\XB)z}}^1\,\frac{\dd u}{u}\,
\int_{z}^1\,\frac{\dd \rho}{\rho}\,A(v(\rho,u))\nonu
&&\cdot\Bigg[
f_{i/\AAA}\left(\frac{\XB}{u}\right)\,
D_{\hh/i}\left(\frac{z}{\rho}\right)\,
C_F\,\Phi_{2qq}^L\nonu
&&+
f_{i/\AAA}\left(\frac{\XB}{u}\right)\,
D_{\hh/g}\left(\frac{z}{\rho}\right)\,
C_F\,\Phi_{2qg}^L\nonu
&&+
f_{g/\AAA}\left(\frac{\XB}{u}\right)\,
D_{\hh/i}\left(\frac{z}{\rho}\right)\,
\frac{1}{2}\,\Phi_{2gq}^L
\Bigg]+\porder{\epsilon}
\eeqn
\setlength{\mathindent}{\conserve}
\setlength{\conserve}{\mathindent}
\setlength{\mathindent}{\wideformula}
\beqn
{\cal A}_{C^L}&=&Y^L\,\sum_{i=q,\qb}\,c_i\,\frac{\alpha_s}{2\pi}\,
\int_{\frac{\XB}{1-(1-\XB)z}}^1\,\frac{\dd u}{u}\,A(1)\nonu
&&\cdot\Bigg[
M_{i,h/\AAA}\left(\frac{\XB}{u},(1-\XB)z\right)\,
C_F\,\Phi_{q}^L\nonu
&&+
M_{g,h/\AAA}\left(\frac{\XB}{u},(1-\XB)z\right)\,
\frac{1}{2}\,\Phi_{g}^L
\Bigg]
+\porder{\epsilon}
\eeqn
\setlength{\mathindent}{\conserve}
\noindent
Here the subtracted and unsubtracted Altarelli-Parisi splitting functions are
\cite{16,24}:
\beqn
P_{q\leftarrow q}(u)&=&C_F\left[2\left(\frac{1}{1-u}\right)_+
+\frac{3}{2}\delta(1-u)-1-u\right]\\
P_{q\leftarrow g}(u)&=&\frac{1}{2}\left[1-2u+2u^2
\right]\\
P_{g\leftarrow q}(u)&=&C_F\left[2\frac{1}{u}-2+u
\right]\\
\hat{P}_{gq\leftarrow q}(u)&=&C_F\left[2\frac{1}{1-u}-1-u
\right]\\
\hat{P}_{\qb q\leftarrow g}(u)&=&\frac{1}{2}\left[1-2u+2u^2
\right]
\eeqn

\newpage
\section{Explicit Formulas for the Finite Contributions}
\label{appexplicit}
In this appendix the explicit expressions for the finite contributions are
presented. They are
\setlength{\conserve}{\mathindent}
\setlength{\mathindent}{\wideformula}
\beqn
\Phi^M_{1qq}&=&\delta(1-u)\,\delta(1-\rho)\,\frac{\pi^2}{3}\nonu
&&\!\!\!\!\!+
\delta(1-\rho)\,\left[
2\left(\frac{\ln(1-u)}{1-u}\right)_{+u[0,\underline{1}]}
+1-u-(1+u)\ln(1-u)-\frac{1+u^2}{1-u}\ln\frac{u-\XB}{1-\XB}
                \right]
\nonu
&&\!\!\!\!\!+
\delta(1-u)\,\left[
2\left(\frac{\ln(1-\rho)}{1-\rho}\right)_{+\rho[0,\underline{1}]}
+1-\rho-(1+\rho)\ln(1-\rho)+\frac{1+\rho^2}{1-\rho}\ln\rho
                \right]
\nonu
&&\!\!\!\!\!
+2\left(\frac{1}{1-\rho}\right)_{+\rho[0,\underline{1}]}\,
\left(\frac{1}{1-u}\right)_{+u[0,\underline{1}]}
-\left(\frac{1}{1-u}\right)_{+u[0,\underline{1}]}\,(1+\rho)
-\left(\frac{1}{1-\rho}\right)_{+\rho[0,\underline{1}]}\,(1+u)
\nonu
&&\!\!\!\!\!
+(1-\rho)\frac{\XB}{u-\XB}
\left(1+\frac{u(1-\XB)}{u-\XB}\right)
-2\frac{u\XB}{u-\XB}
+2
\eeqn
\setlength{\mathindent}{\conserve}
\setlength{\conserve}{\mathindent}
\setlength{\mathindent}{\wideformula}
\beqn
\Phi^M_{1qg}&=&\delta(\rho-a(u))\,
\left[1-u+\frac{1+u^2}{1-u}\ln\frac{1-u}{u}
\right]
\nonu
&&\!\!\!\!\!
+\left(\frac{1}{\rho-a(u)}\right)_{+\rho[\underline{a(u)},1]}
\frac{1+u^2}{1-u}
+\frac{(1-\XB)^2}{(u-\XB)^2}\frac{u^2}{1-u}\rho
-\frac{\XB(1-\XB)}{(u-\XB)^2}u
-2\frac{1-\XB}{u-\XB}\frac{u^2}{1-u}
\eeqn
\setlength{\mathindent}{\conserve}
\setlength{\conserve}{\mathindent}
\setlength{\mathindent}{\wideformula}
\beqn
\Phi^M_{1gq}&=&
\delta(\rho-a(u))\,
\left[2u(1-u)+(1-2u+2u^2)\ln\frac{1-u}{u}
\right]
\nonu
&&\!\!\!\!\!
+\delta(1-\rho)\,
\left[2u(1-u)+(1-2u+2u^2)\ln\frac{(1-u)(1-\XB)}{u-\XB}
\right]
\nonu
&&\!\!\!\!\!
+\left(\frac{1}{\rho-a(u)}\right)_{+\rho[\underline{a(u)},1]}
(1-2u+2u^2)
+\left(\frac{1}{1-\rho}\right)_{+\rho[0,\underline{1}]}
(1-2u+2u^2)
-2\frac{1-\XB}{u-\XB}u
\eeqn
\setlength{\mathindent}{\conserve}
\setlength{\conserve}{\mathindent}
\setlength{\mathindent}{\wideformula}
\beqn
\Phi^M_{2qq}&=&\Phi^M_{1qq}
\eeqn
\setlength{\mathindent}{\conserve}
\setlength{\conserve}{\mathindent}
\setlength{\mathindent}{\wideformula}
\beqn
\Phi^M_{2qg}&=&\delta(1-u)\,
\left[\rho+(\ln \rho+\ln(1-\rho))\left(\rho+\frac{2}{\rho}-2\right)
\right]
\nonu
&&\!\!\!\!\!
+\left(\frac{1}{1-u}\right)_{+u[0,\underline{1}]}
\left(\rho+\frac{2}{\rho}-2\right)
\nonu
&&\!\!\!\!\!
+2-2\frac{\XB u}{u-\XB}
+\frac{\XB}{u-\XB}\rho
-\frac{\XB(1-\XB)u}{(u-\XB)^2}(1-\rho)
+\frac{1}{\rho-a(u)}\frac{1+u^2}{1-u}
-2\frac{1}{\rho}\frac{1}{1-u}
\eeqn
\setlength{\mathindent}{\conserve}
\setlength{\conserve}{\mathindent}
\setlength{\mathindent}{\wideformula}
\beqn
\Phi^M_{2gq}&=&
\delta(1-\rho)\,
\left[2u(1-u)+(1-2u+2u^2)\ln\frac{(1-u)(1-\XB)}{u-\XB}
\right]
\nonu
&&\!\!\!\!\!
+\frac{1}{\rho-a(u)}
(1-2u+2u^2)
+\left(\frac{1}{1-\rho}\right)_{+\rho[0,\underline{1}]}
(1-2u+2u^2)
-2\frac{1-\XB}{u-\XB}u
\eeqn
\setlength{\mathindent}{\conserve}
\setlength{\conserve}{\mathindent}
\setlength{\mathindent}{\wideformula}
\beqn
\Phi^M_{q}&=&(1-\XB)\Bigg[
\frac{7}{2}\,\delta(1-u)\,
-\frac{3}{2}\,\left(\frac{1}{1-u}\right)_{+u[0,\underline{1}]}
+2\,\left(\frac{\ln(1-u)}{1-u}\right)_{+u[0,\underline{1}]}
\nonu
&&\!\!\!\!\!
\quad\quad
\quad\quad
+3-u-(1+u)\ln(1-u)-\frac{1+u^2}{1-u}\ln u
\Bigg]
\eeqn
\setlength{\mathindent}{\conserve}
\setlength{\conserve}{\mathindent}
\setlength{\mathindent}{\wideformula}
\beqn
\Phi^M_{g}&=&(1-\XB)
\left(\ln\frac{1-u}{u}-1\right)(1-2u+2u^2)
\eeqn
\setlength{\mathindent}{\conserve}
\setlength{\conserve}{\mathindent}
\setlength{\mathindent}{\wideformula}
\beqn
\Phi^L_{1qq}&=&
2\frac{(1-\XB)^2}{(u-\XB)^2}u^3\rho
-2\frac{\XB(1-\XB)}{(u-\XB)^2}u^2(1-u)
\eeqn
\setlength{\mathindent}{\conserve}
\setlength{\conserve}{\mathindent}
\setlength{\mathindent}{\wideformula}
\beqn
\Phi^L_{1qg}&=&
2\frac{(1-\XB)^2}{(u-\XB)^2}u^3(1-\rho)
\eeqn
\setlength{\mathindent}{\conserve}
\setlength{\conserve}{\mathindent}
\setlength{\mathindent}{\wideformula}
\beqn
\Phi^L_{1gq}&=&
4\frac{1-\XB}{u-\XB}u^2(1-u)
\eeqn
\setlength{\mathindent}{\conserve}
\setlength{\conserve}{\mathindent}
\setlength{\mathindent}{\wideformula}
\beqn
\Phi^L_{2qq}&=&\Phi^L_{1qq}
\eeqn
\setlength{\mathindent}{\conserve}
\setlength{\conserve}{\mathindent}
\setlength{\mathindent}{\wideformula}
\beqn
\Phi^L_{2qg}&=&\Phi^L_{1qg}
\eeqn
\setlength{\mathindent}{\conserve}
\setlength{\conserve}{\mathindent}
\setlength{\mathindent}{\wideformula}
\beqn
\Phi^L_{2gq}&=&\Phi^L_{1gq}
\eeqn
\setlength{\mathindent}{\conserve}
\setlength{\conserve}{\mathindent}
\setlength{\mathindent}{\wideformula}
\beqn
\Phi^L_{q}&=&(1-\XB)\,u
\eeqn
\setlength{\mathindent}{\conserve}
\setlength{\conserve}{\mathindent}
\setlength{\mathindent}{\wideformula}
\beqn
\Phi^L_{g}&=&(1-\XB)\,2u(1-u)
\eeqn
\setlength{\mathindent}{\conserve}
\end{appendix}

\newpage
%
% bibliography
%
\newcommand{\bibitema}[1]{\bibitem[#1]{#1}}

\newpage
\noindent
{\Large \bf Figure Captions:}
 
\begin{description}
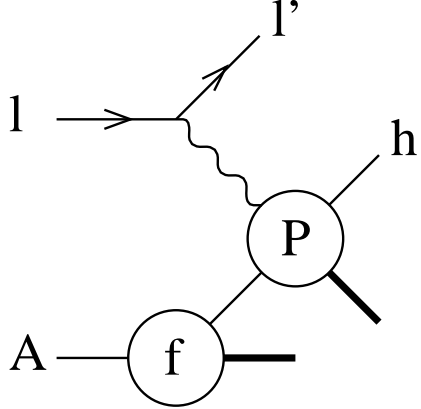
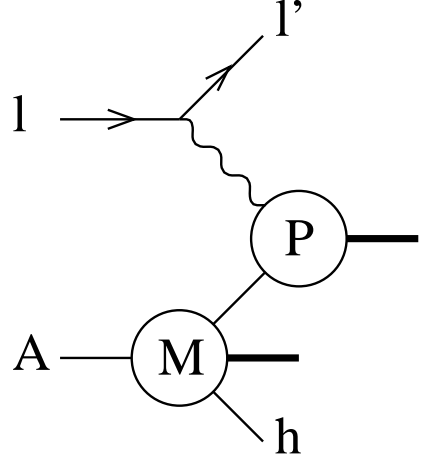
\item[Fig.~1] Contributions to the one-particle inclusive cross section:
current (a), target (b). $f$ stands for the parton densities,
$M$ for fracture functions
and $P$ for any hard process.
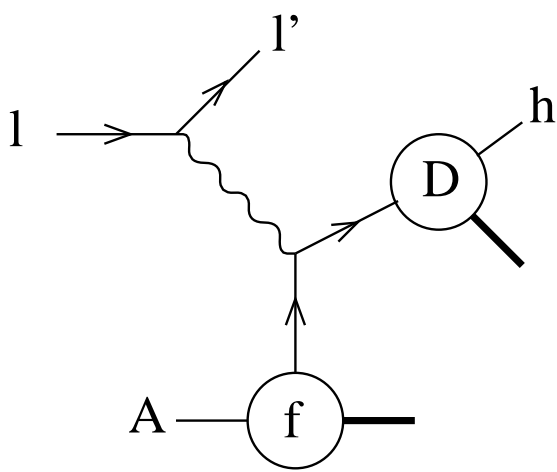
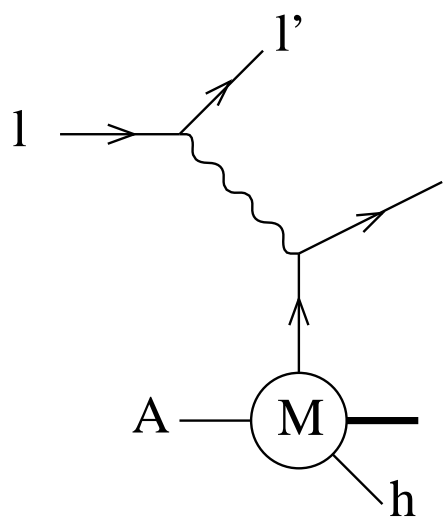
\item[Fig.~2] The parton-model process for the current contribution (a)
and target contribution (b). $D$ is the fragmentation function.
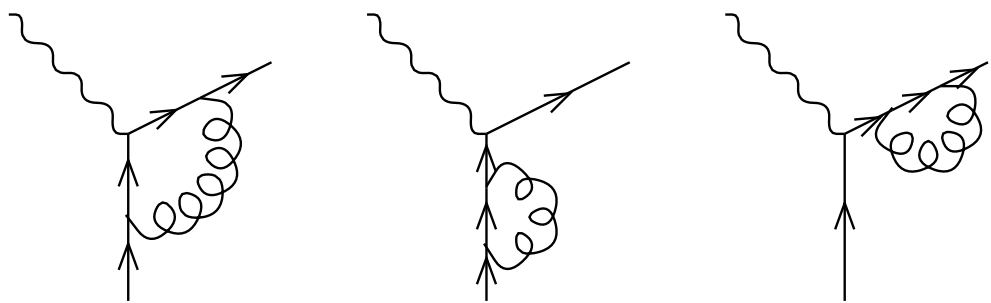
\item[Fig.~3] The virtual corrections to \porder{\alpha_s}.
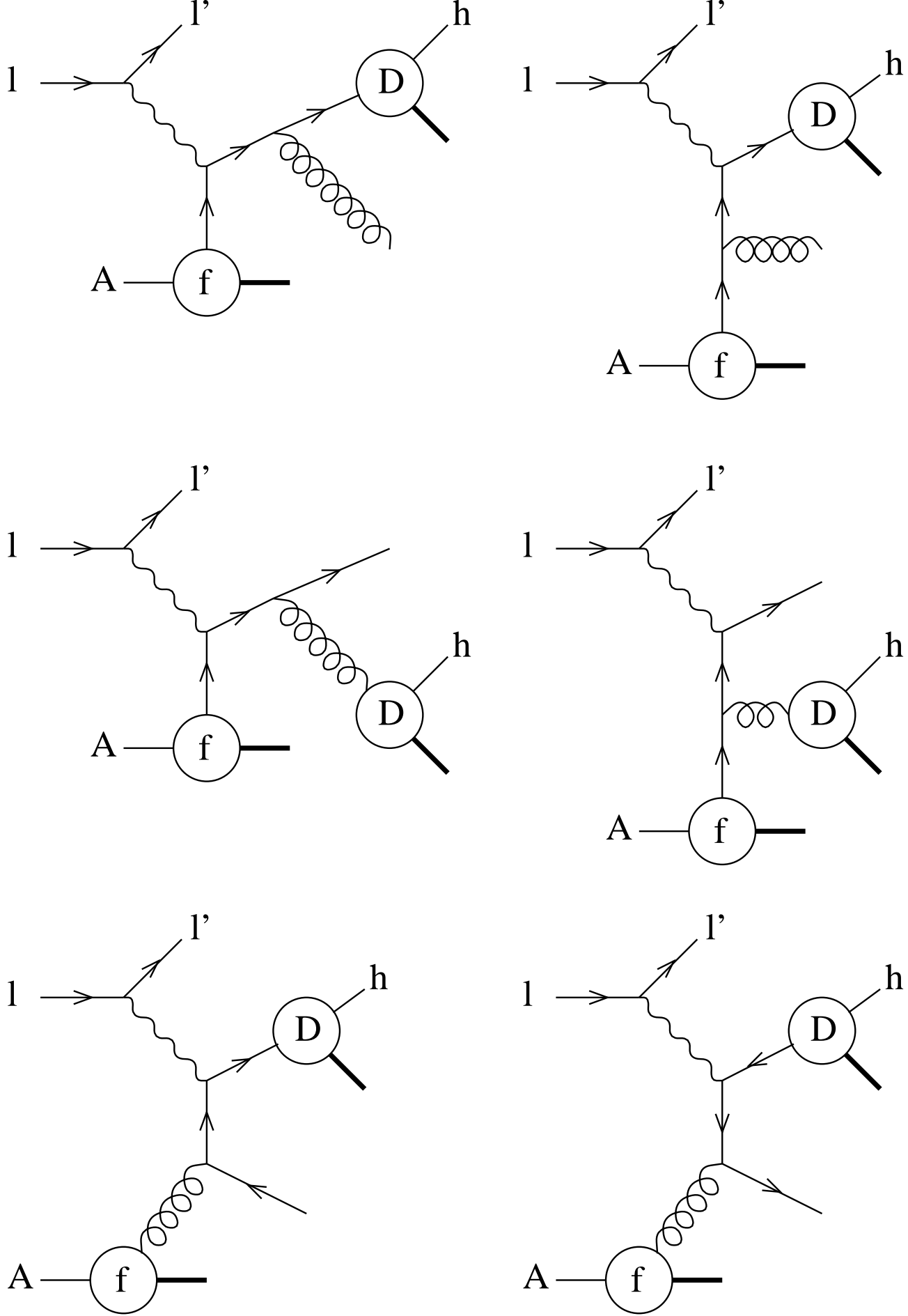
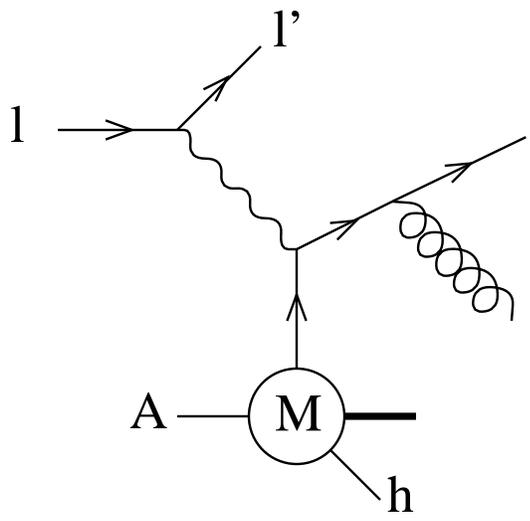
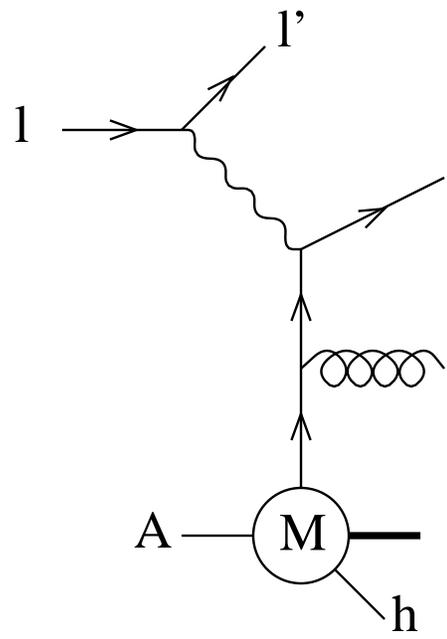
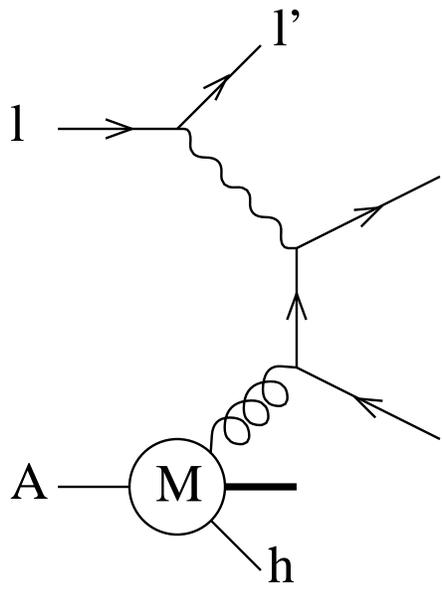
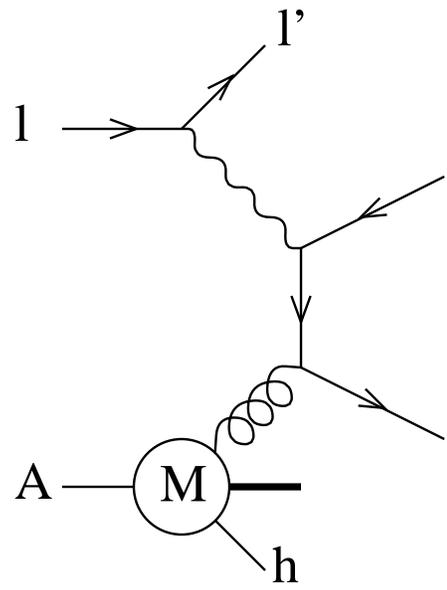
\item[Fig.~4] The real corrections to \porder{\alpha_s}:
diagrams contributing to the terms 
$B_\alpha^A$ (a) and $C^A$ (b).
\end{description}

\end{document}